\documentclass[reprint,superscriptaddress,nofootinbib,amsmath,amssymb,aps,prd,amsfonts,showkeys]{revtex4-2}

\usepackage{graphicx}
\usepackage{dcolumn}
\usepackage{bm}
\usepackage{hyperref}

\usepackage{graphicx}
\usepackage{float}
\usepackage{subfloat}
\usepackage[caption=false]{subfig} 
\usepackage{subfig}
\usepackage{blindtext, subfig}
\usepackage{ragged2e} 
\DeclareCaptionJustification{justified}{\justifying}
\usepackage{textcomp}
\usepackage{amsfonts}
\usepackage{graphics}
\usepackage{epstopdf}
\usepackage{slashed}
\usepackage{multirow}
\usepackage{adjustbox}
\usepackage{lipsum}
\usepackage{rotating}
\usepackage{xcolor}
\usepackage{wrapfig}
\usepackage{epsfig}
\usepackage{siunitx,adjustbox}
\usepackage[compat=1.1.0]{tikz-feynman}
\usepackage[justification=raggedright, margin=5mm,font=sf,labelfont=bf, format=plain]{caption}
\definecolor{rosso}{cmyk}{0,1,1,0.4}
\definecolor{rossos}{cmyk}{0,1,1,0.55}
\definecolor{rossoc}{cmyk}{0,1,1,0.2}
\definecolor{blu}{cmyk}{1,1,0,0.3}
\definecolor{blus}{cmyk}{1,1,0,0.6}
\definecolor{bluc}{cmyk}{1,1,0,0.1}
\definecolor{verde}{cmyk}{0.92,0,0.59,0.25}
\definecolor{verdec}{cmyk}{0.92,0,0.59,0.15}
\definecolor{verdes}{cmyk}{0.92,0,0.59,0.4}

\DeclareRobustCommand{\orcidicon}{\hspace{-2.1mm}
\begin{tikzpicture}
\draw[lime,fill=lime] (0,0.0) circle [radius=0.13] node[white] {{\fontfamily{qag}\selectfont \tiny ID}}; \draw[white,fill=white] (-0.0525,0.095) circle [radius=0.007];
\end{tikzpicture} \hspace{-3.7mm} }
\foreach \x in {A, ..., Z} {\expandafter\xdef\csname orcid\x\endcsname{\noexpand\href{https://orcid.org/\csname orcidauthor\x\endcsname} {\noexpand\orcidicon}}}

\begin{document}
\title{The scale invariant scotogenic model: CDF-II $W$-boson mass and the 95~GeV excesses}

\author{Amine Ahriche\orcidA}
\email{ahriche@sharjah.ac.ae}
\affiliation{Department of Applied Physics and Astronomy, University of Sharjah, P.O. Box 27272 Sharjah, UAE.}

\author{Mohamed Lamine Bellilet}
\email{medlamine.bellilet@outlook.com}
\affiliation{Laboratoire de Physique des Rayonnements, Badji Mokhtar University, B. P. 12, 23000 Annaba, Algeria.}

\author{Mohammed Omer Khojali\orcidK}
\email{khogali11@gmail.com}
\affiliation{School of Physics and Institute for Collider Particle Physics, University of the Witwatersrand, Johannesburg, Wits 2050, South Africa.}
\affiliation{Department of Physics, University of Khartoum, PO Box 321, Khartoum 11115, Sudan.}

\author{Mukesh Kumar\orcidM}
\email{mukesh.kumar@cern.ch}
\affiliation{School of Physics and Institute for Collider Particle Physics, University of the Witwatersrand, Johannesburg, Wits 2050, South Africa.}

\author{Anza-Tshildzi Mulaudzi}
\email{anza-tshilidzi.mulaudzi@cern.ch}
\affiliation{School of Physics and Institute for Collider Particle Physics, University of the Witwatersrand, Johannesburg, Wits 2050, South Africa.}

\begin{abstract}
The anomalies observed in the $W$ mass measurements at the CDF-II experiments and the excesses seen around 95~GeV at the Large Hadron Collider (LHC) motivate this work, in which we investigate and constrain the parameter space of the Scale Invariant Scotogenic Model with a Majorana dark matter candidate. The scanned parameters are chosen to be consistent with the dark matter relic density and the observed excesses at $\sim95$~GeV signal strength rates in different channels. We found that significant part of the viable space addresses the excess in the channel $\gamma\gamma$, while a tight part can address the excess in both $\gamma\gamma$ and $b\bar{b}$ channels. Furthermore, the model's viable parameters can be probed in both the LHC and future $e^{+}e^{-}$ colliders for di-Higgs production.
\end{abstract}

\maketitle

\section{Introduction}

\label{intro}

In the Standard Model (SM), the mass of the $W$-boson is a fundamental parameter, and precise measurements of this mass are crucial for testing the model's predictions. The reported measurements at CDF-II~\cite{CDF:2022hxs} have shown a significant discrepancy between the measured $W$-boson mass ($M_{W}^{{\rm CDF}}=80.4335\pm0.0094$~GeV) and the mass predicted by the SM $m_{W}=80.357\pm0.006$~GeV~\cite{ParticleDataGroup:2022pth}. This discrepancy is said to be at the level of 7 standard deviations. The $W$-boson is a weak interaction carrier; and any deviation from its SM-predicted properties, including its mass, has important implications, potentially indicating the presence of new physics beyond the Standard
Model (BSM). Note that a recent measurement from ATLAS~\cite{ATLAS:2023fsi} ($M_{W}^{{\rm ATLAS}}=80.370\pm0.019$ GeV) shows no deviation from the SM expectation. Excluding the recent measurements from CDF-II~\cite{CDF:2022hxs}, the current world average from experiments yields $M_{W}^{{\rm avg.}}=80.377\pm0.012$~GeV, based on measurements at LEP-2~\cite{ALEPH:2013dgf}, Tevatron~\cite{CDF:2012gpf,D0:2013jba}, and the LHC~\cite{ATLAS:2017rzl,LHCb:2021bjt}.

In search of a light scalar Higgs boson, the CMS and ATLAS experiments at the Large Hadron Collider (LHC) reported a local excess of 2.9$\sigma$ and 1.7$\sigma$ at 95.4~GeV in the di-photon ($\gamma\gamma$) invariant mass spectrum in Run~2 dataset~\cite{CMS:2018cyk,Biekotter:2023oen,CMS:2023yay,ATLAS:2023jzc}, respectively. The Higgs boson ($H$) production in the Higgsstrahlung process $e^{+}e^{-}\to ZH$ with $H\to b\bar{b}$ an excess of 2.3$\sigma$ has been observed in the mass range 95~GeV $<m_{H}<$ 100~GeV at the LEP collider experiments~\cite{LEPWorkingGroupforHiggsbosonsearches:2003ing,OPAL:2002ifx}. CMS also reported another local excess in the light-Higgs boson searches in the $\tau^{+}\tau^{-}$ final state with a significance of 3.1$\sigma$ which is compatible with the aforementioned excesses~\cite{CMS:2022goy}. A recent study estimates the global significance of the excesses at 95~GeV to be 3.8$\sigma$~\cite{Bhattacharya:2023lmu}.

The notable discovery of the Higgs boson at the LHC~\cite{ATLAS:2012yve,CMS:2012qbp} marks the completion of the SM's foundation. Nevertheless, the observed anomalies mentioned above open new avenues for considering and constraining BSM physics. Several such studies are being considered in Refs.~\cite{Cacciapaglia:2016tlr,Crivellin:2017upt,Cao:2019ofo,Biekotter:2019kde,Cline:2019okt,Abdelalim:2020xfk,Heinemeyer:2021msz,Biekotter:2021qbc,Biekotter:2021ovi,Li:2022etb,Iguro:2022dok,Biekotter:2022jyr,Benbrik:2022azi,Biekotter:2022abc,Botella:2022rte,Escribano:2023hxj,Borah:2023hqw,Coloretti:2023wng,Bhattacharya:2023lmu,Abouabid:2023mbu,Ashanujjaman:2023etj,Ahriche:2023wkj}.

Despite its success, the SM has left many questions unanswered, including the hierarchy problem, the nature of dark matter (DM), and the smallness of neutrino masses. Among the extensions of the SM that address these three problems simultaneously is the Scale Invariant Scotogenic Model (SI-SCM)~\cite{Ahriche:2016cio}. In this framework, the SM is extended by a real scalar singlet, three Majorana singlet fermions and an inert scalar doublet. The real scalar singlet develops a vacuum expectation value (VEV) to assist in the radiatively induced electroweak symmetry breaking (EWSB), à la Coleman~\cite{Coleman:1973jx}. Here, we have two CP-even scalars whose tree-level eigenmasses are 0 and 125~GeV which correspond to a dilaton and a SM-like Higgs, respectively. When considering the radiative corrections (RCs), two scenarios are possible: (1) the dilaton mass squared acquires a positive nonzero value, $m_{D}<m_{H}$, and the Higgs mass remains $m_{H}=125$ GeV (light dilaton case); and (2) the zero mass value shifts to 125~GeV due to the RCs, and the 125~GeV tree-level eigenstate becomes a heavy scalar, with $m_{S}>m_{H}$, referred as the Pure Radiative Higgs Mass (PRHM) case~\cite{Ahriche:2021frb}.

In this setup, the new Yukawa interactions that couple the Majorana singlet fermions and the inert scalar doublet to the lepton doublets induce a neutrino mass at one-loop level similar to the minimal scotogenic model~\cite{Ma:2006km}. The DM candidate here could be either a scalar (the lightest neutral inert scalar), resembling the case of the inert Higgs model extended by a real scalar~\cite{Khojali:2022squ}, or the lightest Majorana singlet fermion~\cite{Soualah:2021xbn}. The Majorana DM scenario in this model differs from the minimal scotogenic model, as DM annihilation occurs additionally into all SM fermions and gauge bosons via processes mediated by the Higgs boson and dilaton. This makes the new Yukawa coupling restricted only by the requirements of neutrino oscillation data and lepton flavour constraints. Here, in the setup, we investigate whether the dilaton scalar field could address the 95~GeV excess mentioned previously while considering theoretical and experimental constraints and requirements, including the DM relic density and direct detection, and the $W$-boson mass values measured by CDF-II. In addition, we would like to investigate the impact of all these assumptions and constraints on the di-Higgs production at the LHC (and at future $e^{+}e^{-}$ colliders) with $\sqrt{s}=14$~TeV (500 GeV).

This work is organized as follows: Section~\ref{sec:The-Model} is dedicated to presenting the SI-SCM model, describing EWSB, and discussing various theoretical and experimental constraints. Next, in Section~\ref{sec:MW}, we delve into the discussion and formulation of $W$ mass corrections and the 95~GeV signal strength modifiers in the SI-SCM model. The di-Higgs production mechanism is detailed in Section~\ref{sec:2H}, and our numerical results are presented and discussed in Section~\ref{sec:NR}. We conclude our work in Section~\ref{sec:Conc}.
\begin{table}[t]
\begin{centering}
\begin{tabular}{ccccc}
\hline
\textbf{{Gauge group}}  & $S$  & $N_{i}$  & $\phi$  & $X_{\rm SM}$ \tabularnewline
\hline
$SU(2)_{L}$  & 2  & 1  & 1  & \tabularnewline
\hline
$U(1)_{Y}$  & -1  & 0  & 0  & \tabularnewline
\hline
$Z_{2}$  & -1  & -1  & 1  & 1 \tabularnewline
\hline
 &  &  &  & \tabularnewline
\end{tabular}
\end{centering}
\caption{The field charges under the symmetry $Z_{2}$, where $X_{\rm SM}$ denotes all SM fields.}
\label{charge}
\end{table}

\begin{figure}[t]
\begin{centering}
\includegraphics[width=0.4\textwidth]{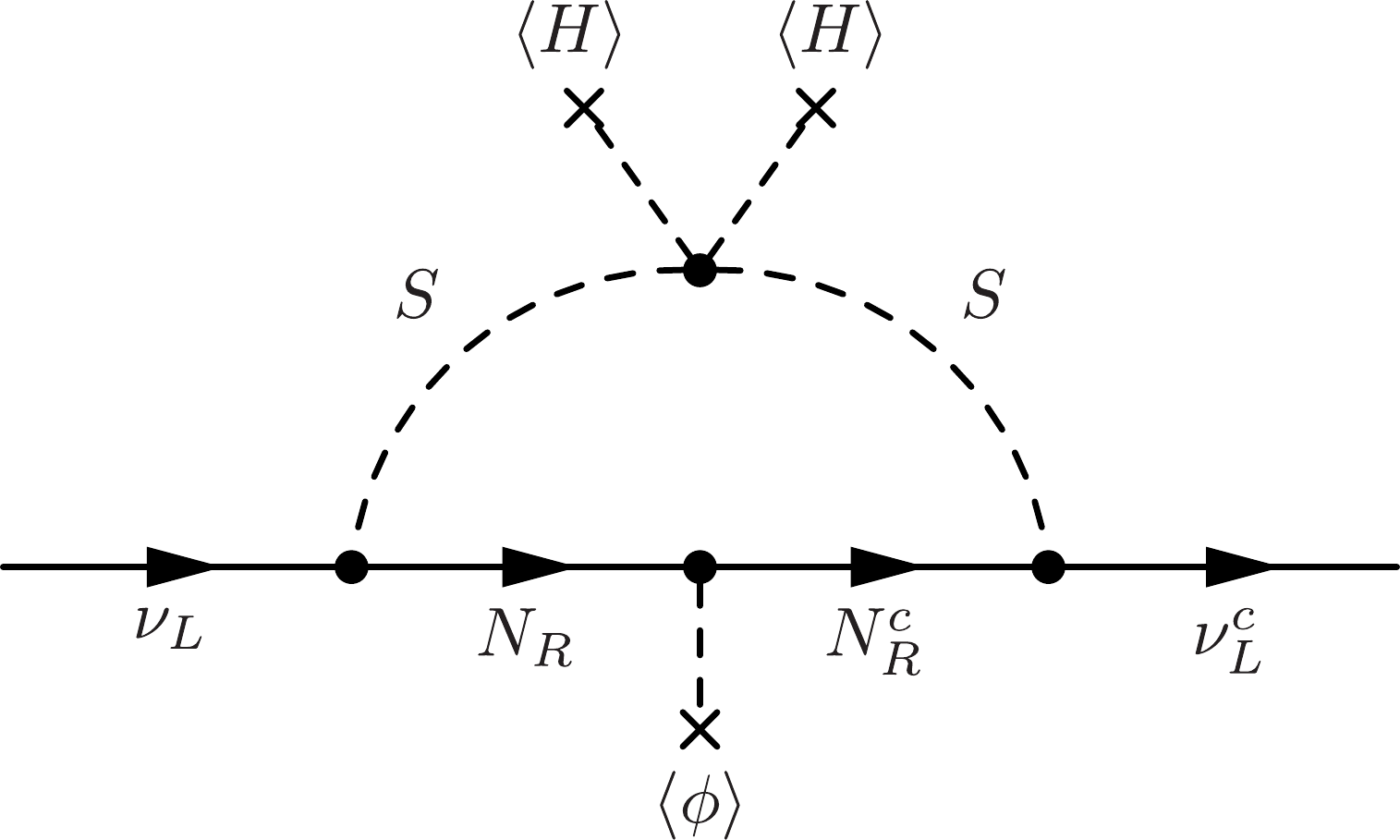}
\end{centering}
\caption{The neutrino mass is generated in the SI-scotogenic model at one-loop level.}
\label{fig:SI3l}
\end{figure}

\section{Model and Framework}

\label{sec:The-Model}

In the SI-SCM, the SM is extended by one inert doublet scalar, $S$, three singlet Majorana fermions $N_{i}$ ($i=1,2,3$), and one real singlet scalar $\phi$ to assist the radiative EWSB, as shown in Table~\ref{charge}. The model is assigned by a global $Z_{2}$ to make the lightest $Z_{2}$-odd field stable, which plays the DM candidate role. The Lagrangian contains the following terms
\begin{align}
\mathcal{L}\supset & -\;\{g_{i,\alpha}\overline{N_{i}^{c}}S^{\dagger}L_{\beta}+\mathrm{h.c}\}-\frac{1}{2}y_{i}\phi\overline{N_{i}^{c}}\,N_{i}
,\label{L:Ma}
\end{align}
where, $g_{i,\alpha}$ and $y_{i}$ are new Yukawa couplings; $L_{\beta}$ are ($\ell_{\alpha R}$) the left-handed lepton doublet (right-handed leptons); the Greek letters label the SM flavours, $\alpha,\,\beta\in\{e,\,\mu,\,\tau\}$; the SM Higgs and the inert scalar doublets are parameterised as: $\mathcal{H}^{T}=\Big(\chi^{+},(h+i\,\chi^{0})/\sqrt{2}\Big)$ and $S^{T}=\Big(S^{+},(S^{0}+i\,A^{0})/\sqrt{2}\Big)$, respectively (where $\chi^{+}$ and $\chi^{0}$are Goldstone bosons). The most general SI scalar potential that obeys the $Z_{2}$ symmetry is given by
\begin{align}
V(\mathcal{H},&\,S,\,\phi)  =\frac{1}{6}\lambda_{H}(\left|\mathcal{H}\right|^{2})^{2}+\frac{\lambda_{\phi}}{24}\phi^{4}+\frac{\lambda_{S}}{2}|S|^{4}\nonumber \\
 & +\frac{\omega}{2}|\mathcal{H}|^{2}\phi^{2}+\frac{\kappa}{2}\,\phi^{2}|S|^{2}+\lambda_{3}\,|\mathcal{H}|^{2}|S|^{2}\nonumber \\
 & +\lambda_{4}\,|\mathcal{H}^{\dagger}S|^{2}+\left\{ \frac{\lambda_{5}}{2}(\mathcal{H}^{\dagger}S)^{2}+h.c.\right\} ,\label{V:Ma}
\end{align}
The first term in eq.~(\ref{L:Ma}) and the last term in eq.~(\ref{V:Ma}) are responsible for generating neutrino mass via the one-loop diagrams as illustrated in Fig.~\ref{fig:SI3l}.

The neutrino mass matrix element~\cite{Merle:2015gea} can be written as $m_{\alpha\beta}^{(\nu)}=\sum_{i}g_{i,\alpha}g_{i,\beta}\Lambda_{i}=\left(g^{T}\cdot\varLambda\cdot g\right)_{\alpha\beta}$, which permits us to estimate the new Yukawa couplings using to the Casas-Ibarra parameterization~\cite{Casas:2001sr}, where lepton flavour violating (LFV) bounds on the branching ratios of $\ell_{\alpha}\to\ell_{\beta}\gamma$ and $\ell_{\alpha}\to\ell_{\beta}\ell_{\beta}\ell_{\beta}$ should be fulfilled.

Here, the EWSB is triggered by the RCs where the counter-term $\delta\lambda_{H},\,\delta\lambda_{\phi},\,\delta\omega$ corresponding to terms in eq.~(\ref{V:Ma}), are chosen to fulfil the tadpole conditions and one of the CP-even eigenmasses matches the 125 SM-like Higgs and the other corresponds to light Higgs (PRHM case) or a heavy Higgs (light dilaton case). After the EWSB ($\langle h\rangle=\upsilon,\,\langle\phi\rangle=x$), we obtain two CP-even eigenstates as $H=c_{\alpha}~h-s_{\alpha}~\phi$ and $D=s_{\alpha}~h+c_{\alpha}~\phi$, where $H$ denotes the 125 $\mathrm{GeV}$ Higgs, $D$ is the dilaton scalar whose mass should be around $m_{D}=95.4\,\mathrm{GeV}$ in this setup; and $\alpha$ is the Higgs-dilaton mixing angle. Here, the RCs in both PRHM and light dilaton cases ensure the mixing angle $\alpha$ to be in the experimental range dictated by the Higgs gauge couplings measurements. Detailed discussions on these conditions can be found in~\cite{Ahriche:2021frb}.

The vacuum stability must be ensured by imposing the coefficients of the term $\phi^{4}\log\phi$ to be positive, which represents the leading term in the scalar; instead the term $\phi^{4}$, where $\phi$ refers to any direction in the $h-\phi$ plane. Since all field dependent squared masses can be written as $m_{i}^{2}(h,\phi)=\frac{1}{2}(\alpha_{i}h^{2}+\beta_{i}\phi^{2})$, the vacuum stability conditions can be written as $\sum_{i}n_{i}\alpha_{i}^{2}>0$ and $\sum_{i}n_{i}\beta_{i}^{2}>0$, with $n_{i}$ to be the multiplicity of the field ``$i$". In addition to these conditions, the quartic couplings in eq.~(\ref{V:Ma}) must fulfil the perturbative unitarity conditions~\cite{Ahriche:2021frb}.

In this model, the DM candidate could be fermionic (the lightest Majorana fermion, $N_{1}$) or a scalar (the lightest among $S^{0}$ and $A^{0}$). In the case of a scalar DM, the situation matches the singlet extended inert doublet model case~\cite{Khojali:2022squ}, where the co-annihilation effect should be considered in order to have viable parameters space. In the minimal scotogenic model with Majorana DM, the DM annihilation occurs via $t$-channel diagrams mediated by the inert fields, which makes the Yukawa couplings $g_{i,\alpha}$ values constrained by the relic density, and therefore the neutrino mass smallness can be achieved only in extreme $S^{0}-A^{0}$ mass degeneracy, i.e., imposing a very small value for $\lambda_{5}\sim\mathcal{O}(10^{-10})$~\cite{Ahriche:2020pwq}. However, in the scale-invariant version, new $s$-channels mediated by the Higgs-boson or dilaton exist, which allows all the perturbative range for the $g_{i,\alpha}$ Yukawa couplings. Also, it is worth noting that, in contrast to many Majorana dark matter models, in this model, the dark matter couples to quarks at the tree level. This feature underscores the significance of direct detection constraints on the parameter space~\cite{Soualah:2021xbn}.

\section{$M_{W}$ measurements \& 95~GeV excesses \label{sec:MW}}

The mass of the $W$-boson can be calculated as a function of the oblique parameters $\Delta S$, $\Delta T$, and $\Delta U$, and is given by:
\begin{align}
M_{W} =&\, m_{W}\,\Bigg[1+\frac{\alpha}{c_{W}^{2}-s_{W}^{2}} \times \nonumber \\
&\Big(-\frac{1}{2}\Delta S+c_{W}^{2}\Delta T+\frac{c_{W}^{2}-s_{W}^{2}}{4s_{W}^{2}}\Delta U\Big)\Bigg]^{\frac{1}{2}},\label{eq:mw}
\end{align}
where $c_{W}=\cos\theta_{W}$ and $s_{W}=\sin\theta_{W}$, with $\theta_{W}$ being the weak mixing angle. The oblique parameters in SI-SCM model are given by ~\cite{Grimus:2008nb}
\begin{align}
&\varDelta S  =\frac{1}{24\pi}\Big\{ \left(2s_{W}^{2}-1\right)^{2}G\left(m_{S^{\pm}}^{2},m_{S^{\pm}}^{2},m_{Z}^{2}\right)\nonumber \\
& +G\left(m_{S^{0}}^{2},m_{A^{0}}^{2},m_{Z}^{2}\right)
 +\log\left(\frac{m_{S^{0}}^{2}m_{A^{0}}^{2}}{m_{S^{\pm}}^{4}}\right)\nonumber \\
 &+s_{\alpha}^{2}\left[\log\frac{m_{D}^{2}}{m_{H}^{2}}-\hat{G}\left(m_{H}^{2},m_{Z}^{2}\right)+\hat{G}\left(m_{D}^{2},m_{Z}^{2}\right)\right]\Big\},
\end{align}
\begin{align}
&\varDelta T =\frac{1}{16\pi s_{W}^{2}m_{W}^{2}} \times \nonumber \\
 &\Big\{ F\left(m_{S^{\pm}}^{2},m_{S^{0}}^{2}\right)+F\left(m_{S^{\pm}}^{2},m_{A^{0}}^{2}\right) -F\left(m_{S^{0}}^{2},m_{A^{0}}^{2}\right) \nonumber \\
 & +3s_{\alpha}^{2}\left[F\left(m_{W}^{2},m_{H}^{2}\right)-F\left(m_{Z}^{2},m_{H}^{2}\right)-F\left(m_{W}^{2},m_{D}^{2}\right) \right. \nonumber\\
& +F\left(m_{Z}^{2},m_{D}^{2}\right)]\Big\},
\end{align}
\begin{align}
&\varDelta U =\frac{1}{24\pi}\Big\{ G\left(m_{S^{\pm}}^{2},m_{S^{0}}^{2},m_{W}^{2}\right)+G\left(m_{S^{\pm}}^{2},m_{A^{0}}^{2},m_{W}^{2}\right)\nonumber \\
 & -\left[2s_{W}^{2}-1\right]^{2}G\left(m_{S^{\pm}}^{2},m_{S^{\pm}}^{2},m_{Z}^{2}\right)-G\left(m_{S^{0}}^{2},m_{A^{0}}^{2},m_{Z}^{2}\right)\nonumber \\
 & +s_{a}^{2}\Big[\hat{G}\left(m_{D}^{2},m_{W}^{2}\right)-\hat{G}\left(m_{D}^{2},m_{Z}^{2}\right)-\hat{G}\left(m_{H}^{2},m_{W}^{2}\right) \nonumber\\
 & +\hat{G}\left(m_{H}^{2},m_{Z}^{2}\right)\Big]\Big\} ,
\end{align}
where the one-loop functions $G,\,F$ and $\hat{G}$ can be found in~\cite{Grimus:2008nb}.\footnote{Note: Subsequent to the CDF-II results, several research groups have adjusted their fits for the oblique parameters $\Delta S$, $\Delta T$, and $\Delta U$ in the context of electroweak precision measurements~\cite{CentellesChulia:2022vpz,Flacher:2008zq,Asadi:2022xiy}, examining their potential effects on BSM physics.} The oblique parameter $\varDelta T$ quantifies the contribution of new physics at low energies and $\varDelta S$ at different energy scales.

In order to analyze whether the SI-SCM model can yield a shift in the prediction for $M_{W}$ that is compatible with the measurements at experiments and simultaneously provides a possible explanation of the observed excesses, namely: (1) $\gamma\gamma$ \& $b\bar{b}$ or (2) $\gamma\gamma$, $b\bar{b}$ \& $\tau^{+}\tau^{-}$, we perform a $\chi^{2}$ analysis. This analysis quantifies the agreement between the theoretically predicted signal rates $\mu_{X\bar{X}}$ (where $X=\gamma,\,b$ for case (1) and $X=\gamma,\,b,\,\tau$ for case (2)) and the experimentally observed values $\mu_{X\bar{X}}^{{\rm exp}}$. Experimentally, it was determined that the excesses at $\sim95$~GeV were best described assuming signal rates of a scalar resonance as~\cite{Biekotter:2023oen,OPAL:2002ifx,CMS:2022goy}:
\begin{align}
\left.\begin{array}{l}
\mu_{\gamma\gamma}^{\mathrm{exp}}=0.27_{-0.09}^{+0.10},\\
\\
\mu_{b\bar{b}}^{\mathrm{exp}}=0.117\pm0.057,\\
\\
\mu_{\tau\tau}^{\mathrm{exp}}=1.2\pm0.5
\end{array}\right\} ,\label{eq:muXX}
\end{align}
where the signal strengths are defined as the cross-section times the branching ratios divided by the corresponding predictions for the hypothetical SM Higgs boson at the same mass and the experimental uncertainties are given as 1$\sigma$ uncertainties.

The theoretically predicted values for $\mu_{X\bar{X}}$ can be expressed as
\begin{align}
\mu_{X\bar{X}} &= \frac{\sigma(gg\to D)\cdot\mathcal{B}(D\to X\bar{X})}{\sigma^{{\rm SM}}(gg\to H)\cdot\mathcal{B}^{{\rm SM}}(H\to X\bar{X})} \nonumber \\
&= \rho_{X}\big(1-{\cal B}(D\to X_{{\rm BSM}})\big), \label{eq:mu}
\end{align}
where $\rho_{X}$ is defined for $X = \gamma, b, \tau$ as:
\begin{align}
\rho_{\gamma} &= \left|1+\frac{\upsilon}{2}\frac{\lambda_{DS^{\pm}S^{\mp}}}{m_{S^{+}}^{2}}\frac{A_{0}^{\gamma\gamma}\Big(\frac{m_{D}^{2}}{4m_{S^{+}}^{2}}\Big)}{A_{1}^{\gamma\gamma}\Big(\frac{m_{D}^{2}}{4m_{W}^{2}}\Big)+\frac{4}{3}A_{1/2}^{\gamma\gamma}\Big(\frac{m_{D}^{2}}{4m_{t}^{2}}\Big)}\right|^{2}, \nonumber \\
\rho_{b} &= \rho_{\tau} = s_{\alpha}^{2},
\end{align}
with $X_{{\rm BSM}}=N_{i}N_{k},S^{0}S^{0},A^{0}A^{0}$. Here, $\sigma(gg\to D)$ and $\mathcal{B}(D\to X\bar{X})$ represent the ggF dilaton production and the final state $X\bar{X}$ branching ratio, respectively. The scalar triple coupling of the dilaton with charged scalars, $\lambda_{DS^{\pm}S^{\mp}}$, is given by $\lambda_{DS^{\pm}S^{\mp}}=s_{\alpha}\lambda_{3}\upsilon+c_{\alpha}\kappa x$, and the loop functions $A_{0,1,1/2}^{\gamma\gamma}$ are provided in~\cite{Djouadi:2005gi}. Additionally, we have $\Gamma_{{\rm tot}}^{D}=s_{\alpha}^{2}\Gamma_{{\rm tot}}^{D,\,{\rm SM}}+\Gamma(D\to X_{{\rm BSM}})$, and $\sigma^{{\rm SM}}(gg\to H)$ and $\mathcal{B}^{{\rm SM}}(H\to X\bar{X})$ are the corresponding SM quantities evaluated at the Higgs-boson mass $m_{h}\to m_{D}$~\cite{HevayHiggs}.

To assess the combined description of the three excesses, we define a total $\chi_{(N)}^{2}$ ($N = 2, 3$) function as
\begin{align}
\left.\begin{array}{l}
\chi_{(2)}^{2}=\chi_{\gamma\gamma}^{2}+\chi_{b\bar{b}}^{2}, \\ \\
\chi_{(3)}^{2}=\chi_{\gamma\gamma}^{2}+\chi_{b\bar{b}}^{2}+\chi_{\tau\tau}^{2},
\end{array}\right\}\label{eq:chi2}
\end{align}
and
\begin{align}
\chi_{i}^{2}=\left(\frac{\mu_{i}-\mu_{i}^{\mathrm{exp}}}{\Delta\mu_{i}^{\mathrm{exp}}}\right)^{2},
\end{align}
where $i = \gamma\gamma$, $b\bar{b}$, or $\tau\tau$. These functions are useful for checking whether the excess can be addressed simultaneously in the channels: (i) $\gamma\gamma,\,b\bar{b}$ and (ii) $\gamma\gamma,\,b\bar{b},\,\tau\tau$.

In the parameter space region corresponding to a small charged scalar effect on the effective coupling $D\gamma\gamma$ (i.e., $\rho_{\gamma}\sim1$), as indicated by the experimental values in eq.~(\ref{eq:muXX}), we found the following results for $\chi^2_N < 1\sigma$:
\begin{enumerate}
    \item[(i)] $\chi_{{\rm (2)}}^{2}|_{\rm min}=2.262$: $s_{\alpha}^{2}=0.11$, and ${\cal B}(D\to X_{{\rm BSM}})=70.15\%$.
    \item[(ii)] $\chi_{{\rm (3)}}^{2}|_{\rm min}=7.708$: $s_{\alpha}^{2}=0.11$, and ${\cal B}(D\to X_{{\rm BSM}})=69.7\%$.
\end{enumerate}
This scenario becomes plausible only if the channel $D\to{\rm inv.}$ is accessible. Here, `inv.' denotes invisible channels such as $N_{i}N_{k}, H^{0}H^{0}, A^{0}A^{0}$. Given that the inert neutral masses are expected to be large, the Majorana singlet fermions should be light, i.e., $m_{\rm DM}=M_{1}<m_{D}/2$, as will be confirmed later. In the subsequent numerical analysis (Section-\ref{sec:NR}), we will consider parameter points to provide a good description of the excesses if they account for the combined effect of the excess in the channels $\gamma\gamma$ and $b\bar{b}$ at the $1\sigma$ level, since the $\tau\tau$ channel appears to be hopeless.

\section{The di-Higgs production\label{sec:2H}}

In the SM, the measurement of di-Higgs ($HH$) production is intriguing not only because it enables the determination of Higgs-boson self-interaction but also because it contributes to understanding EWSB. In cases where EWSB proceeds via a single scalar (the SM Higgs boson), the di-Higgs signal occurs through two Feynman diagrams: box and triangle diagrams. The triangle diagram involves a triple Higgs boson vertex that could be modified by new physics effects, represented as $\lambda_{hhh}=\lambda_{hhh}^{\rm(SM)}(1+\Delta_{hhh})$. Therefore, the di-Higgs measurement can precisely determine the new physics effect by measuring $\Delta_{hhh}$. However, when EWSB involves more than one scalar, as in the model considered in this study, the di-Higgs signal occurs through a box and two (or more) triangle diagrams that involve additional triple couplings $\lambda_{hhS}$ (with $S=h, D$) and new CP scalar (dilaton in our case). Consequently, the determination of $\lambda_{hhh}$ is not straightforward since $\sigma(HH)$ depends on several model parameters. Nonetheless, the experimental bounds on $\sigma(HH)$ through different channels, either via resonant or non-resonant production, are very useful for constraining the scalar sector, especially if $\sigma(HH)$ exhibits values greater than the SM predictions. Non-resonant $HH$ production at the LHC occurs primarily through the dominant gluon fusion (ggF) mode and the sub-dominant vector-boson fusion (VBF) mode. The cross section for $HH$ production at next-to-next-to-leading order (NNLO), including finite top-quark-mass effects in the ggF mode,
is $\sigma_{{\rm ggF}}^{{\rm SM}}=31.05_{-7.2}^{+2.1}$ fb~\cite{Dawson:1998py,Borowka:2016ehy,Baglio:2018lrj,deFlorian:2013jea,Shao:2013bz,deFlorian:2015moa,Grazzini:2018bsd,Baglio:2020wgt}. However, in the VBF mode, at next-to-next-to-next-to-leading order (N$^{3}$LO), the cross section is $\sigma_{{\rm VBF}}^{SM}=1.73\pm0.04$ fb~\cite{Baglio:2012np,Frederix:2014hta,Ling:2014sne,Dreyer:2018rfu,Dreyer:2018qbw} for $m_{h}=125$ GeV at $\sqrt{s}=13$ TeV. The smallness of the $HH$ production cross section in ggF mode at leading order (LO) results from the negative interference between the box and triangle Feynman diagrams, determined by three contributing factors~\cite{Ahriche:2014cpa,Baouche:2021wwa,Ahriche:2021frb}:
\begin{equation}
\sigma^{{\rm SM}}(HH)=\sigma_{B}+\sigma_{T}+\sigma_{BT},
\end{equation}
where $\sigma_{B}=70.1$~fb represents the box contribution, $\sigma_{T}=9.66$~fb corresponds to the triangle contribution, and $\sigma_{BT}=-49.9$~fb accounts for their interference~\cite{Spira:1995mt}.

In the SI-SCM framework, non-resonant $HH$ production through ggF mode includes an additional triangle Feynman diagram mediated through the dilaton field $D$. The production cross section of di-Higgs production either at the LHC or at $e^{-}e^{+}$ colliders has three distinct contributions that come from: (1) the Feynman diagrams involving only the triple scalar couplings that have one scalar propagator, (2) the diagrams with only pure gauge couplings; and (3) the interference contribution~\cite{Ahriche:2021frb,Ahriche:2014cpa}. Therefore, the di-Higgs production cross-section can be expressed as follows:
\begin{equation}
\sigma(HH)=\zeta_{1}\sigma_{B}+\zeta_{2}\sigma_{T}+\zeta_{3}\sigma_{BT},\label{XS}
\end{equation}
where the coefficients $\zeta_{i}$ in this model are modified with respect to the SM as~\cite{Ahriche:2014cpa}
\begin{align}
\left.\begin{array}{l}
\zeta_{1}=c_{\alpha}^{4},\\
\\
\zeta_{2}=\left|c_{\alpha}\frac{\lambda_{HHH}}{\lambda_{HHH}^{\text{SM}}}+s_{\alpha}\frac{\lambda_{HHD}}{\lambda_{HHH}^{\text{SM}}}\frac{s-m_{H}^{2}+im_{H}\Gamma_{H}}{s-m_{D}^{2}+im_{D}\Gamma_{D}}\right|^{2},\\
\\
\zeta_{3}=c_{\alpha}^{2}\Re\left(c_{\alpha}\frac{\lambda_{HHH}}{\lambda_{HHH}^{\text{SM}}}+s_{\alpha}\frac{\lambda_{HHD}}{\lambda_{HHH}^{\text{SM}}}\frac{s-m_{H}^{2}+im_{H}\Gamma_{H}}{s-m_{D}^{2}+im_{D}\Gamma_{D}}\right)
\end{array}\right\} ,\label{xis}
\end{align}
where $\lambda_{HHH}^{\text{SM}}$ is the Higgs triple coupling in the SM; $\sqrt{s}$ is the center-of-mass collision energy, which we will consider to be $\sqrt{s}=14$~TeV at LHC. The expression for the one-loop triple Higgs coupling in the SM is~\cite{Kanemura:2004mg}:
\begin{equation}
\lambda_{HHH}^{\text{SM}}\simeq\frac{3m_{H}^{2}}{\upsilon}\left[1-\frac{m_{t}^{4}}{\pi^{2}\upsilon^{2}m_{H}^{2}}\right],
\end{equation}
with $m_{t}$ is the top quark mass.

Interestingly, a direct measurement of the triple Higgs-boson self-coupling is achievable through resonant $HH$ production at a future $e^{+}e^{-}$ collider. This involves double Higgsstrahlung processes with $W$ or $Z$ bosons, as well as through $WW$ or $ZZ$ fusion. In the case of double Higgsstrahlung ($e^{+}e^{-}\to HHZ$) production at $\sqrt{s}=500$~GeV, the production cross-section can be expressed as eq.~(\ref{XS}) using the same coefficients in eq.~(\ref{xis}) and the cross-section contributions given as $\sigma_{B}=0.0837$ fb, $\sigma_{T}=0.01565$~fb, and $\sigma_{BT}=0.05685$~fb~\cite{Ahriche:2021frb}.

\section{Numerical Results and Discussion\label{sec:NR}}

In this section, our attention is directed towards the parameter space corresponding to the dilaton mass window of 94~GeV $<m_{D}<$ 97~GeV. We systematically consider various theoretical and experimental constraints, including perturbativity, perturbative unitarity, Higgs boson di-photon and invisible decays, LEP negative searches, and the electroweak precision tests. In addition, we require the DM relic density to match the observed values, and the DM direct detection constraints to be satisfied~\cite{Soualah:2021xbn}. In our analysis, instead of relying on the bounds on the total Higgs strength modifier ($\mu_{{\rm tot}}=c_{\alpha}^{2}\times(1-\mathcal{B}_{{\rm inv.}})\geq0.89$ at 95\% confidence level (C.L.)~\cite{ATLAS:2016neq})\footnote{Here, the Higgs invisible branching ratio is constrained by ATLAS to be $\mathcal{B}_{{\rm inv.}}=\mathcal{B}(H\to N_{i}N_{k})<0.11$~\cite{ATLAS:2020kdi}.}, we perform a detailed analysis by considering the bounds on the Higgs total decay width ($\Gamma_{h}=4.6_{-2.5}^{+2.6}~{\rm MeV}$; and the partial Higgs strength signal modifiers $\mu_{XX}^{h}$ for $X=\mu,\tau,b,\gamma,W,Z$~\cite{ParticleDataGroup:2022pth}. For this reason, we define the SM $\chi_{{\rm SM}}^{2}$ function
\begin{equation}
\chi_{{\rm SM}}^{2}=\sum_{\mathcal{O}}\chi_{\mathcal{O}}^{2}=\sum_{\mathcal{O}}\left(\frac{\mathcal{O}-\mathcal{O}^{\mathrm{exp}}}{\Delta\mathcal{O}^{\mathrm{exp}}}\right)^{2},\label{eq:chi2-SM}
\end{equation}
with the observables $\mathcal{O}$ denotes the Higgs total decay width ($\Gamma_{h}$) and the Higgs signal strength modifiers ($\mu_{XX}^{h}$). In our analysis, we consider only benchmark points (BPs) with a precision of 95\% C.L., i.e., $\chi_{{\rm SM}}^{2}<11.07$.

Additionally, we ensure that the considered values for the model's free parameters, including the inert masses ($m_{S^{0}}$, $m_{A^{0}}$, $m_{S^{\pm}}$), Majorana masses ($M_{i}$), scalar coupling $\lambda_{3}$, and the singlet VEV $x$, correspond to values of the new Yukawa couplings $g_{i,\alpha}$ that satisfy the neutrino oscillation data and LFV constraints.

\begin{figure*}[t]
\subfloat[\label{fig:PSa}]{\includegraphics[width=0.45\textwidth]{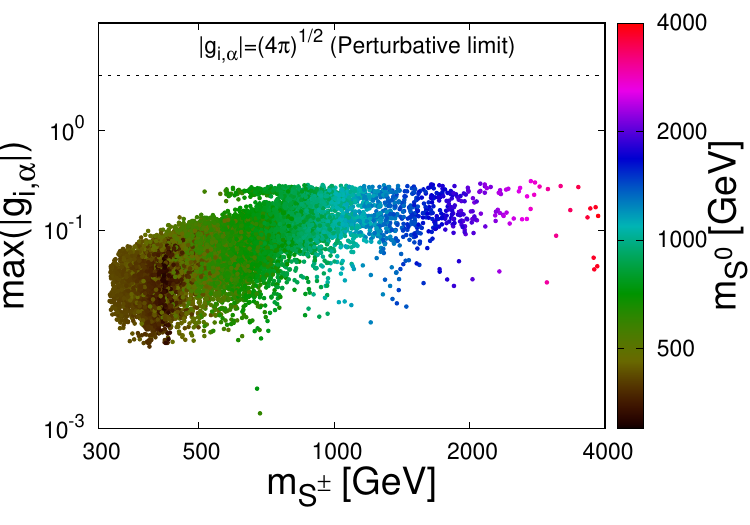}}\qquad
\subfloat[\label{fig:PSb}]{\includegraphics[width=0.45\textwidth]{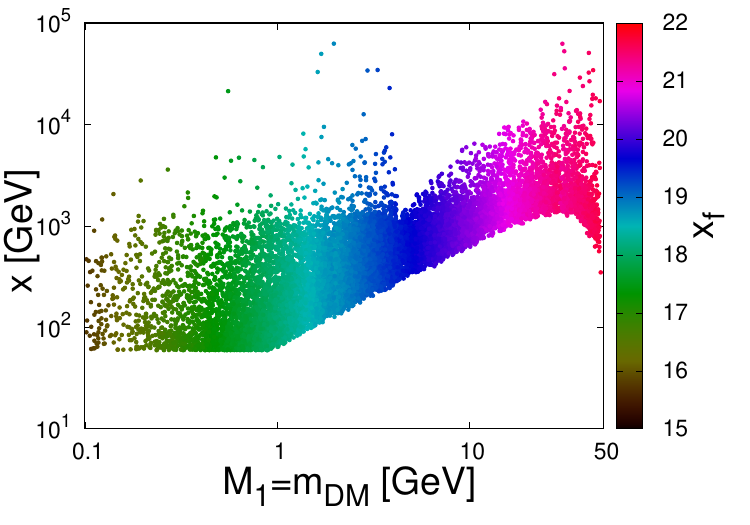}}\\
\subfloat[\label{fig:PSc}]{\includegraphics[width=0.65\textwidth]{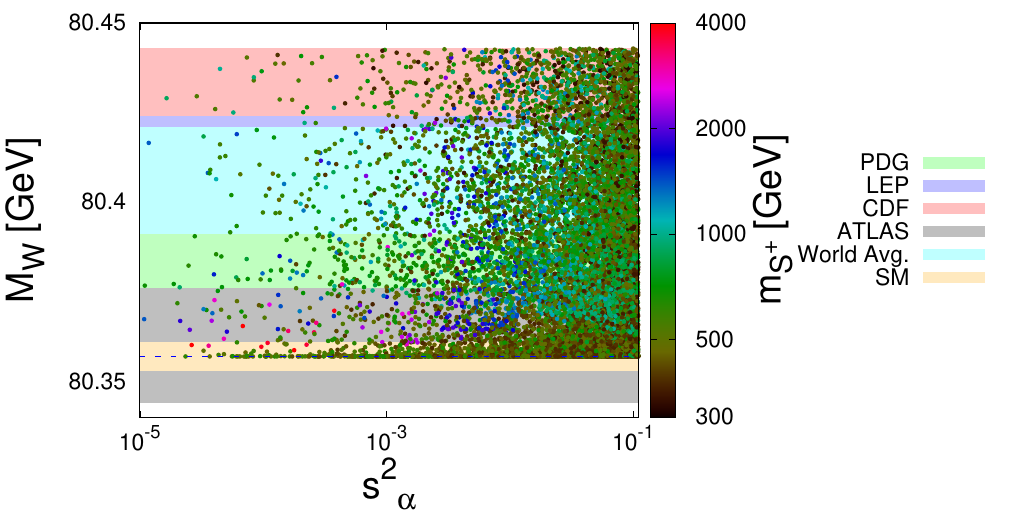}}
\caption{(a) Maximum Yukawa couplings $\max(|g_{i,\alpha}|)$ plotted against the masses of the charged inert doublet, $m_{S^{\pm}}$, and the neutral scalar, $m_{S^{0}}$. (b) The singlet scalar VEV $x$ as a function of $m_{{\rm DM}}$ and the freeze-out parameter, $x_{f}$. (c) $M_{W}$ prediction in the SI-SCM model shown with respect to the scalar mixing angle, $s_{\alpha}^{2}$, and $m_{S^{\pm}}$. The horizontal color bands represent $M_{W}$ measurements at a $2\sigma$ level from various experiments: green for PDG~\cite{Lu:2022bgw}, blue for LEP~\cite{L3:2005fft}, red for CDF~\cite{CDF:2022hxs}, grey for ATLAS~\cite{ATLAS:2023fsi}, cyan for the world average~\cite{Ashanujjaman:2023etj}, and yellow for the SM value.}
\label{fig:PS}
\end{figure*}

Through a random numerical scan adhering to the aforementioned constraints and conditions, we consider 15k BPs that satisfy the 95~GeV excess in the $\gamma\gamma$ channel ($\mu^{\rm exp}_{\gamma\gamma}=0.27^{+0.10}_{-0.09}$). The viable parameter space fulfilling the various conditions and constraints is illustrated in Fig.~\ref{fig:PS}.

\begin{figure*}[t]
\centering
\subfloat[\label{fig:chi2a}]{\includegraphics[width=0.48\textwidth]{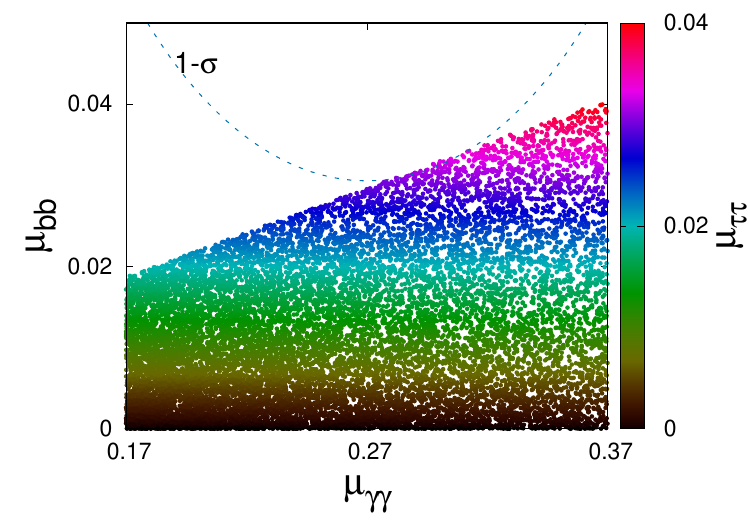}} \quad
\subfloat[\label{fig:chi2b}]{\includegraphics[width=0.48\textwidth]{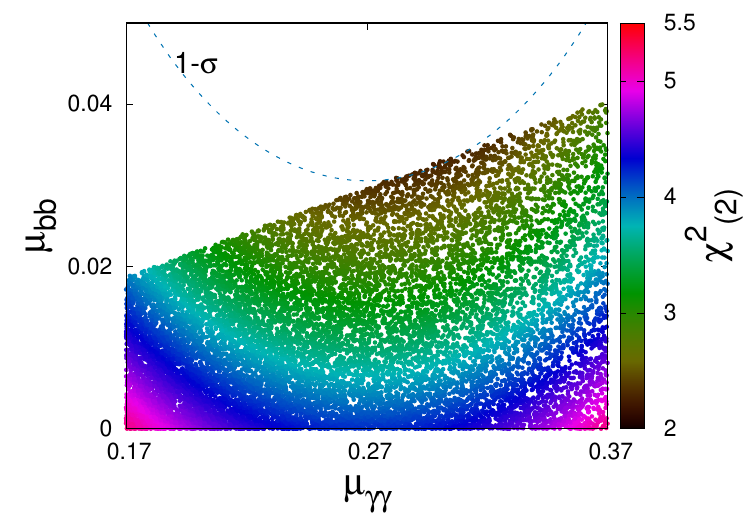}}\\
\subfloat[\label{fig:chi2c}]{\includegraphics[width=0.48\textwidth]{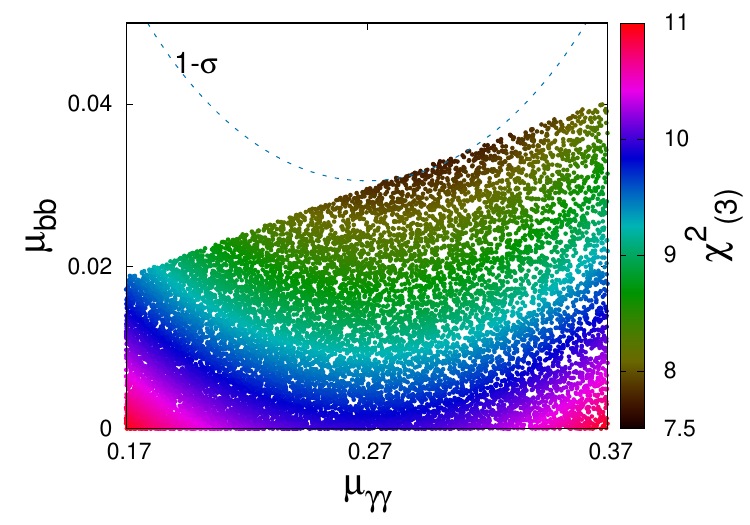}}
\quad
\subfloat[\label{fig:chi2d}]{\includegraphics[width=0.48\textwidth]{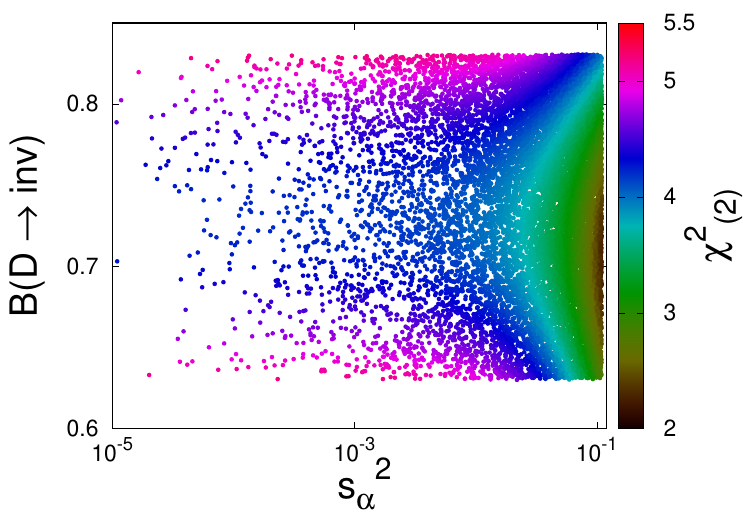}}
\caption{(a) The plot illustrates the signal strengths for the three excesses: $\gamma\gamma$, $\tau\tau$, and $b\bar{b}$. (b) This plot presents the signal strengths of $\gamma\gamma$ and $b\bar{b}$ as functions of the total $\chi_{(2)}^{2}$. (c) Signal strengths of $\gamma\gamma$ and $b\bar{b}$ are plotted against the total $\chi_{(3)}^{2}$. (d) This plot shows the dilaton's invisible branching ratio versus the scalar mixing angle, $s_{\alpha}^{2}$, presented alongside the total $\chi_{(2)}^{2}$. In panels (a), (b), and (c), all the benchmark points (BPs) fall within the $2\sigma$ contour, though it is visible only in a restricted region due to the chosen $\mu_{\gamma\gamma}$ values.}
\label{fig:chi2}
\end{figure*}
\begin{figure*}[t]
\centering
\subfloat[\label{fig:HHa}]{\includegraphics[width=0.48\textwidth]{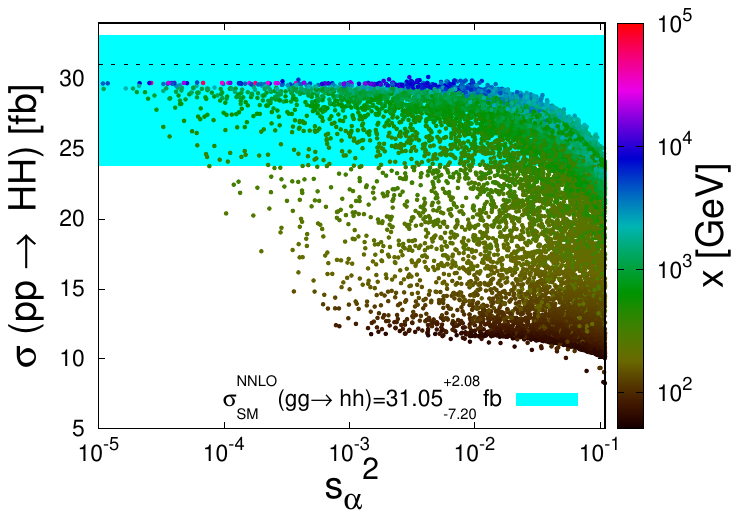}}
\quad
\subfloat[\label{fig:HHb}]{\includegraphics[width=0.48\textwidth]{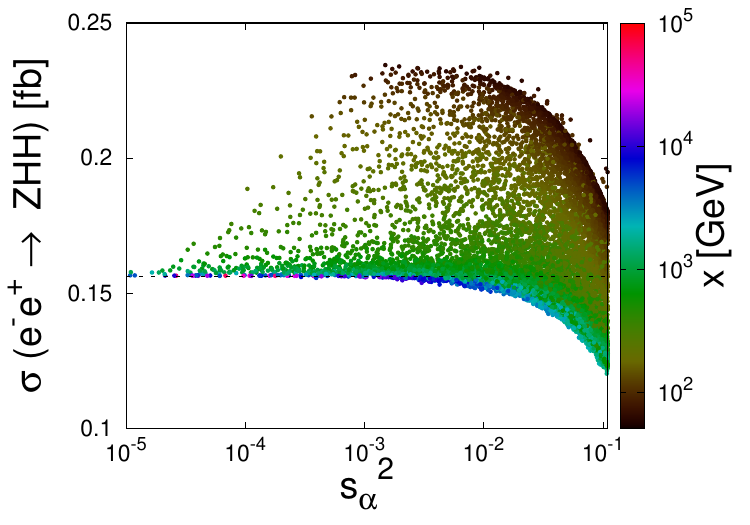}}
\caption{The di-Higgs production cross-sections: (a) through ggF mode at the LHC with $\sqrt{s}=14$~TeV and (b) at future $e^{-}e^{+}$ colliders with $\sqrt{s}=500$~GeV, plotted against the scalar mixing angle $s_{\alpha}^{2}$. The color palette represents the singlet VEV $x$ in GeV. The black dashed lines indicate the SM predictions.
}
\label{fig:HH}
\end{figure*}

Fig.~\ref{fig:PSa} reveals that the assumptions of $m_{D}\approx95$~GeV and $\mu^{\rm exp}_{\gamma\gamma}=0.27^{+0.10}_{-0.09}$ lead to inert masses exceeding 300~GeV. Additionally, the new Yukawa couplings are an order of magnitude smaller than the perturbative limit, contrasting with general cases in the SI-SCM~\cite{Ahriche:2016cio,Soualah:2021xbn}. From Fig.~\ref{fig:PSa}, one can observe that the couplings $g_{i,\alpha}$ have values ranging from 0.001 to 0.4 for inert masses across different ranges. These values are dictated by the requirements for DM relic density~\cite{Ahriche:2017iar,Ahriche:2020pwq,Soualah:2021xbn}, where the DM annihilation channels $N_1N_1 \to \ell_{\alpha}\ell_{\beta},\nu_{\alpha}\nu_{\beta}$ play a key role in the majority of the viable parameter space.

In Fig.~\ref{fig:PSb}, we show the singlet scalar VEV $x$ as a function of the DM mass $m_{\rm DM}$ and the freez-out parameter $x_f$. It is noteworthy that $m_{\rm DM}$ must be smaller than $m_{D}/2$ to ensure the branching ratio ${\cal B}(D\to\text{inv.})$ lies between 50\% and 85\%, satisfying the assumption $\mu^{\rm exp}_{\gamma\gamma}=0.27^{+0.10}_{-0.09}$ (Fig.~\ref{fig:PSb}). It's important to mention that the values of the DM freeze-out observable (the annihilation cross section and freeze-out parameter $x_{f}=m_{\rm DM}/T_{f}$, where $T_{f}$ is the freeze-out temperature) exhibit typical values for a Weakly Interacting Massive Particle DM candidate. One notices that the singlet scalar VEV, $x$, has lower bounds arising from various constraints. Perturbativity and unitarity together requires $x>57$~GeV, which is the dominant constraint for $M_1\equiv m_{\rm DM}\lesssim 1$~GeV. For $M_1\equiv m_{\rm DM}>1$~GeV, the dominant constraint comes from the Higgs-dilaton mixing, which is equivalent to $\chi^2_{\rm SM} < 11.07$. Despite the tight range of the dilaton mass, 94~GeV $< m_{D} <$ 97~GeV, the singlet scalar VEV spans a broad interval, $x \in [57~{\rm GeV}, 100~{\rm TeV}]$. This can be understood by considering that the dilaton mass is a one-loop effect sensitive to multiple parameters, including $x$, the multiplicity ($n_i$), and the couplings of the new fields to the Higgs doublet ($\alpha_i$) and the singlet scalar ($\beta_i$). Thus, any value for the dilaton mass can be achieved by selecting appropriate values for the couplings ($\alpha_i$, $\beta_i$), making the range of $x$ indistinguishable for cases of restricted and non-restricted dilaton mass~\cite{Ahriche:2016cio, Ahriche:2021frb, Soualah:2021xbn}.

In Fig.~\ref{fig:PSc}, we depict the $M_{W}$ prediction within the framework of the SI-SCM model as a function of the scalar mixing angle $s_{\alpha}^{2}$ and $m_{S^{\pm}}$. Additionally, we compare these predictions with various measurements at their respective 2$\sigma$ limits. Importantly, in this model, the correction $\varDelta m_{W}$ is strictly positive, driven by the fact that $\varDelta T$ is always positive and dominates over the values of $\varDelta S$ and $\varDelta U$. For instance, without considering the $M_{W}$ bounds, the relative mass difference between CP-even and CP-odd scalars could reach 20\%, whereas incorporating the $M_{W}$ measurements reduces this difference to below 5\%.

In Fig.~\ref{fig:chi2}, we present the signal strength modifiers and the relevant observable for the $95$ GeV scalar candidate. Notably, the excesses $\mu_{\gamma\gamma,~b\bar{b}}$ can be simultaneously addressed, whereas the excess $\mu_{\tau\tau}$ shows suppressed values. As illustrated in the $\mu_{\gamma\gamma}-\mu_{b\bar{b}}$ panels in Fig.~\ref{fig:chi2}, the excesses in the channels ($\gamma\gamma,~b\bar{b}$) can be accommodated simultaneously at the $1\sigma$ level for a very small region of the parameter space (the BPs point inside the $1\sigma$ contour in Fig.~\ref{fig:chi2a}, \ref{fig:chi2b} and~\ref{fig:chi2c} with $\chi^2_{(2)}<2.53$); and at the 2$\sigma$ level for a significant region of the parameter space. However, the $\tau\tau$ channel can only be accommodated at the 99\% C.L., i.e., $8.02<\chi_{(3)}^{2}<11.34$. The preference for maximal values in both scalar mixing, $s_{\alpha}^{2}\approx0.11$, and ${\cal B}(D\to{\rm inv.})\approx70\%$, becomes evident when matching the excess in channels other than $\gamma\gamma$ as shown in Fig.~\ref{fig:chi2d}. If the di-tau excess is analyzed by ATLAS and/or re-analyzed by CMS with additional data, and the measured $\mu_{\tau\tau}^{{\rm exp}}$ is relaxed to around 0.6-0.7, then addressing the three excesses within this model becomes feasible.

In Fig.~\ref{fig:HH}, we present the di-Higgs production cross section for both (a) the LHC with $\sqrt{s}=14$~TeV and (b) future $e^{-}e^{+}$ colliders with $\sqrt{s}=500$~GeV plotted against the scalar mixing angle $s_{\alpha}^{2}$. As shown in Fig.~\ref{fig:HHa}, di-Higgs production at the LHC exhibits no enhancement; however, a reduction of 73\% is possible for benchmark points with a smaller singlet VEV $x$ and non-suppressed scalar mixing. In contrast, at $e^{+}e^{-}$ colliders, the double Higgsstrahlung cross-section ranges from a reduction of 24\% to an enhancement of 46\% compared to the SM cross-section, as illustrated in Fig.~\ref{fig:HHb}.

Current constraints from the LHC~\cite{ATLAS:2021ifb, ATLAS:2023gzn, ATLAS:2022hwc,ATLAS:2022xzm,ATLAS:2022fpx,ATLAS:2023qzf,ATLAS:2023elc,ATLAS:2024yuv,ATLAS:2024lsk,ATLAS:2024lhu}, particularly those from the $HH \to \gamma\gamma b\bar{b}$ and other di-Higgs decay channels, place stringent limits on the di-Higgs production cross-section and the Higgs cubic self-coupling. Our model, which predicts a non-zero branching ratio to invisible particles, must be evaluated in light of these constraints. The di-Higgs cross-section predicted by our model is smaller than the SM values, inherently satisfying all the bounds from negative searches at the LHC. Given the smaller cross-section, the predicted rates for $HH \to \gamma\gamma b\bar{b}$ and other di-Higgs decay channels remain below the upper limits set by the LHC. This implies that the model does not predict an excess of di-Higgs events that would have already been excluded by current experimental data.

While our results suggest that the SI-SCM model is consistent with current studies on collider anomalies, including the CDF-II $W$-mass anomaly and the 95~GeV excess~\cite{Escribano:2023hxj,Borah:2023hqw}, it's important to note that the muon's magnetic moment, $(g-2)_{\mu}$, anomaly cannot be addressed in this model. This limitation arises from its single negative contribution, which is insufficient to match the measured value. To reconcile this discrepancy in $(g-2)_{\mu}$, an extension of the SI-SCM by incorporating additional scalar components may be required. Such an extension could potentially enable the model to successfully account for the experimental values associated with $(g-2)_{\mu}$.

\section{Conclusion\label{sec:Conc}}

In response to anomalies observed in the measurement of the $W$-mass at CDF-II and an excess around $\sim95$~GeV reported by LEP, CMS, and ATLAS, we conducted a study within the framework of the SI-SCM model to address these issues. This model, characterized by radiatively induced EWSB, not only accommodates light neutrino masses but also proposes a Majorana dark matter candidate. Its predictions are well within the reach of collider experiments. We identified a viable parameter space where inert scalar masses can explain the $W$-mass anomaly, and the light dilaton with $m_{D}\sim95$~GeV aligns with the observations in that region. Under these assumptions, the dilaton might need to decay into invisible Majorana singlet fermions to address the excess (\ref{eq:muXX}) in the channels $\gamma\gamma~\&~b\bar{b}$. For this reason, the excesses considered in this study can be accommodated simultaneously at the $1\sigma$ level within a very small region of the parameter space, and at the $2\sigma$ level within a significant region. Within this parameter space, di-Higgs production at the LHC shows no enhancement compared to the SM. However, at $e^{+}e^{-}$ colliders, the double Higgsstrahlung cross-section varies from a reduction of 24\% to an enhancement of 46\% compared to the SM cross-section.

\acknowledgements
A.A. and M.L.B. were funded by the University of Sharjah under the research projects No 21021430107 ``\textit{Hunting for New Physics at Colliders}" and No 23021430135 ``\textit{Terascale Physics: Colliders vs Cosmology}."


\end{document}